# Reconsidering the gold open access citation advantage postulate in a multidisciplinary context: an analysis of the subject categories in the Web of Science database 2009-2014

Pablo Dorta-González [a,*], Sara M. González-Betancor [b], María Isabel Dorta-González [c]

[a] Universidad de Las Palmas de Gran Canaria, TiDES Research Institute, Campus de Tafira, 35017 Las Palmas de Gran Canaria, Spain. *E-mail*: pablo.dorta@ulpgc.es

[b] Universidad de Las Palmas de Gran Canaria, Departamento de Métodos Cuantitativos en Economía y Gestión, Campus de Tafira, 35017 Las Palmas de Gran Canaria, Spain. *E-mail*: sara.gonzalez@ulpgc.es

[c] Universidad de La Laguna, Departamento de Ingeniería Informática y de Sistemas, Avenida Astrofísico Francisco Sánchez s/n, 38271 La Laguna, Spain. *E-mail*: isadorta@ull.es

* Corresponding author and proofs. E-mail: pablo.dorta@ulpgc.es (P. Dorta-González).

## Abstract

Since Lawrence in 2001 proposed the open access (OA) citation advantage, the potential benefit of OA in relation to the citation impact has been discussed in depth. The methodology to test this postulate ranges from comparing the impact factors of OA journals versus traditional ones, to comparing citations of OA versus non-OA articles published in the same non-OA journals. However, conclusions are not entirely consistent among fields, and two possible explications have been suggested in those fields where a citation advantage has been observed for OA: the early view and the selection bias postulates. In this study, a longitudinal and multidisciplinary analysis of the gold OA citation advantage is developed. All research articles in all journals for all subject categories in the multidisciplinary database Web of Science are considered. A total of 1,137,634 articles – 86,712 OA articles (7.6%) and 1,050,922 non-OA articles (92.4%) – published in 2009 are analysed. The citation window considered goes from 2009 to 2014, and data are aggregated for the 249 disciplines (subject categories). At journal level, we also study the evolution of journal impact factors for OA and non-OA journals in those disciplines whose OA prevalence is higher (top 36 subject categories). As the main conclusion, there is no generalizable gold OA citation advantage, neither at article nor at journal level.

## Introduction

The publication of results obtained during a scientific research is the final stage of a long period involving the planning, execution, and analysis of results. This publication stage has benefited greatly from the emergence of Internet (Björk, 2004). In the Internet age, more researchers are making their research openly accessible to increase the visibility, usage, and citation impact. Open Access (OA) was defined in 2002 by Budapest Open Access Initiative as free and unrestricted access on the public Internet to literature that scholars provide without expectation of direct payment (Prosser, 2003).

There are two modalities of OA (Harnad et al., 2004): gold OA refers to articles in fully accessible OA journals; green OA refers to publishing in a traditional journal, in addition to self-archiving the pre- or post-print paper in a repository. Currently, the Directory of Open Access Journals (DOAJ) is the largest index presenting quality controls of scientific journals that allows open access. According to the DOAJ, in March 2016 there were 4,989 journals that did not require an article processing charge (APC), 2,205 that did, while no information was available about the processing charge of another 2,195 journals.

Many researchers, starting with Lawrence (2001), have found that OA articles tend to have more citations than pay-for-access articles. This citation advantage has been observed in a variety of academic fields including computer science (Lawrence, 2001), physics (Harnad et al., 2004), philosophy, political science, electrical and electronic engineering, and mathematics (Antelman, 2004), biology and chemistry (Eysenbach, 2006), as well as civil engineering (Koler-Povh, Južnič & Turk, 2014).

However, since Lawrence proposed in 2001 the OA citation advantage, this postulate has been discussed in the literature in depth, without achieving an agreement (Davis et al., 2008; Gargouri et al., 2010; Joint, 2009; Norris et al., 2008; Wang et al., 2015). Some authors are critical about the causal link between OA and higher citations, stating that the benefits of open access are uncertain and may vary among different fields (Craig et al., 2007; Davis & Walters, 2011).

Kurtz et al. (2005), and later other authors (Craig et al., 2007; Davis et al., 2008; Moed, 2007), set out three postulates supporting the existence of a correlation between open access and increased citations, concluding that early view and selection bias effects are the main factors behind this correlation:

a) *The Open Access postulate*. Since open access articles are easier to obtain, they are easier to read and cite.

b) *The Early View postulate*. Open access articles tend to be available online prior to their publication. They can therefore begin accumulating citations earlier than paid-access articles published at the same time. When comparing citations at fixed times since publication, the open-access articles will have more citations because they have been available for longer.

c) *The Selection Bias postulate*. If more prominent authors are more likely to provide open access to their articles, or if authors are more likely to provide open access to their highest quality articles, then open access articles will have more citations than paid-access articles.

Niyazov et al. (2016), and Gaule & Maystre (2011) found evidence of selection bias in open access, but they still estimated a statistically significant citation advantage even after controlling for that bias. Regardless of the validity or generality of their conclusions, these studies establish that any analysis must take into account the effect of time and selection bias.

At journal level, Gumpenberger, Ovalle-Perandones & Gorraiz (2013) showed that the impact factor of gold OA journals was increasing, and that one-third of newly launched OA journals were indexed in JCR after three years. However, Björk and Solomon (2012) argued that the distribution model is not related to journal impact. This result has been confirmed by Solomon, Laakso & Björk (2013), concluding that regardless of the distribution model, articles are cited at a similar rate.

In the literature related to open access advantage some specific fields have already been analysed, as stated above. However this paper is the first multidisciplinary study that includes all scientific disciplines, and that analyses this effect at journal level as well as at article level. As the main conclusion, it can be advanced that there is no general citation advantage of gold open access at either level.

Finally, another of the aims in this paper is to contrast the prevalence of the OA articles by scientific disciplines and its changes over time. As a brief summary, the percentage of OA articles has increased in the time period 2009-2014 in all three indexes (60.4% in SCIE, 30.8% in SSCI, and 5.5% in AHCI).

## Methodology

In this study we have analysed exclusively the gold open access, that is, journals in which all the articles that are published are open access. In this sense, those journals that use a hybrid business model that give the possibility of putting articles in open access when the authors pay the APCs, are considered as not-open access journals.

To research whether there is a general citation advantage of gold open access, we restrict our analysis to articles indexed in the Web of Science core collection, 'old enough' to make a robust recounting of its cites. Thus, we considered all research articles published in journals included in the Science Citation Index Expanded (SCIE), the Social Sciences Citation Index (SSCI), and the Arts & Humanities Citation Index (AHCI) during 2009. In order to reduce the early view effect, we consider a citation impact window of six years after publication. Thus the citation time window considered was 2009-2014, meaning that the total number of citations of those articles was measured 6 years after their publication. At the same time, percentages of the OA articles in years 2009 and 2014 were observed in order to contrast their changes over time. Moreover, the journal impact factor (JIF) of all considered journals was observed along the same period 2009-2014.

We perform a double analysis, one article-level and one journal-level. At the article level we aggregate data by scientific disciplines (Web of Science subject categories) and consider the average impact –measured in terms of number of citations– of OA articles and non-OA articles within each subject category. First we perform a descriptive analysis of the total articles, as well as the percentage of OA articles. Then, we compare the average citations of both OA and non-OA articles, considering the ratio between both averages, so that ratios greater than one indicate higher citation averages for OA articles and, conversely, ratios less than one indicate lower citation averages for OA articles. Finally, the relationship between the average citation of OA and non-OA articles is also analysed through a measure of the OA citation advantage. All three analyses are made quantitatively as well as graphically.

As a measure of the OA citation advantage (OACA) we consider the proportion of the average citation of OA articles in relation to non-OA articles. More precisely, denoting by $OAC_i$ the average citation of OA articles in category *i*, and by $NOAC_i$ the average citation of non-OA articles in category *i*, then the OA citation advantage of thematic category *i* can be defined as:

$$OACA_i = \frac{OAC_i - NOAC_i}{NOAC_i} \times 100$$

Therefore, a value of $OACA=p$ means that OA articles are cited *p*% more than non-OA articles. Similarly, a negative value of -*p*% means that OA articles are cited *p*% less than non-OA articles.

At the journal level, we analyse graphically the JIF evolution for the top 36 categories with the highest OA percentages in 2009, to see if there is a common pattern for those categories with a higher prevalence of OA journals that differs from other categories with a lower prevalence of this type of journal.

## Results and discussion

### *Article level analysis*

All the information related to articles has been aggregated by each of the 249 subject categories in which the Web of Science database classified the journals in 2009 (Table 1). There were a total of 1,137,634 articles in 2009. Of those, there were 86,712 (7.6%) articles published in OA journals, that is, in journals in which all published articles are OA. The Web of Science database identifies OA journals and subscription of FECYT (Spanish Foundation for Science and Technology) gives the possibility of filtering by this aspect. In 2014 this figure increased, as there were 1,419,895 articles, of which 183,710 (12.9%) were articles in OA journals.

[Table 1 about here]

In order to analyse the OA prevalence, the total number of research articles and the percentage of OA articles in each subject category are shown in Table 1. We consider both the year of the article's publication (2009) and the end of the citation window (2014).

The smallest categories in the Web of Science, in terms of the number of research articles they include, are 'Literature, African, Australian, Canadian' (with 174 articles in 2009) and 'Poetry' (with 124 articles in 2014). On the other side, the largest categories are 'Biology' (with 84,271 items in 2009), 'Materials Science' and 'Multidisciplinary Sciences' (with 76,382 items in 2014). There is a general increase in the amount of articles in all categories between 2009 and 2014. In fact, and taking as reference points the first and third quartiles of the distribution for the number of articles, 25% of all categories had a maximum of 1,534 articles in 2009, while in 2014 that maximum was 1,995; and 75% of all categories had a maximum of 9,716 articles in 2009, lower than the maximum of 11,041 articles in 2014. Thus, columns four and six in Table 1 (total articles) show that there are important differences in relation to the size among subject categories.

Although there are still categories without OA journals, the amount of such categories has been declining. Thus, in 2009 there were 40 categories without OA journals, while in 2014 there were just 34 categories. On the opposite side, the categories with a higher prevalence of OA articles are 'Tropical Medicine' (with 43.4% of OA in 2009) and 'Multidisciplinary Sciences' (with 73.8% of OA in 2014). Taking again as reference points the first and third quartiles of the distribution for the percentage of OA, there is an increment in the OA prevalence. In fact, 25% of all categories have a maximum of 0.9% of their articles in OA in 2009, which increases to 1.6% in 2014; and 75% of all categories had a maximum OA percentage of 6.9% in 2009, which increases to 11.8% in 2014. Therefore, it is observed that the prevalence of OA articles in each category has also increased over the considered years.

A descriptive analysis of variables total articles and OA prevalence is shown in Table 2. There is a general increase in the scientific production in SCIE and SSCI, but a decrease in AHCI. Thus, the average number of articles published in each category of the SCIE went from near 10,300 to about 11,100, meaning a 7.8% increase. Meanwhile, the average number of articles published in the SSCI categories increased from about 3,400 in 2009 to about 4,200 in 2014, implying an average increase of the scientific production in those categories of 22.4%. By contrast, the average number of articles of all categories in the AHCI went from about 1,560 to about 1,380, implying a decrease of 11.6%.

[Table 2 about here]

The percentage of OA articles in the different categories has increased in the same time period in all three indexes. In SCIE, the average percentage of OA in its different categories went from 6.34% to 10.17%, which implies an increase of 60.4%. In SSCI that percentage went from 3.41% to 4.46%, which implies an increase of 30.8%. Finally, in AHCI that percentage went from 3.85% to 4.06%, implying an increase of 5.5%.

Therefore, it seems that the change in the size of the categories between 2009 and 2014 –understanding by size the amount of articles published in each category– is not related to the change in its percentage of OA articles during the same years. In fact, the categories of the SSCI were the ones that increased more in research articles (22.4%), while the categories with a higher increase in their amount of OA articles were not those of the SSCI, but rather of the SCIE (60.4%).

At the same time, Table 2 shows a high variability between categories within each index. Just looking at the SCIE of 2009 one can see that there are categories with only 112 articles, while others have 84,271 articles. The same happens with the percentage of OA articles in each category. In both considered years all indices have categories with no OA articles, while other categories have as many as 73.8% of their articles in OA (SCIE, 2014).

The increase in the OA prevalence was significant between 2009 and 2014. This conclusion can also be graphically made from Figure 1, since in all three indexes most of the bubbles –each one representing one subject category– are above the bisecting line. In relation to the axes scale, it is relevant to highlight that the OA percentages of many categories of the SCIE are higher than those of the SSCI and the AHCI. The bubble size is proportional to the number of research articles in the category. Regarding this size, it can be clearly seen that there are many more categories in the SCIE than in the SSCI, and within each category there are also many more articles in the SCIE than in the SSCI. A similar relationship occurs between SSCI and AHCI.

[Figure 1 about here]

Average citation in the time window 2009-2014, for articles published in 2009, is shown in Table 1. Most categories show a ratio between OA and non-OA average citations lower than one, from which it follows that in the Web of Science database the articles published in OA journals are generally less cited than those published in pay-for-access journals. Therefore, there is no generalizable OA citation advantage at article level.

This conclusion can also be graphically made from Figure 2, where one can see that most of the bubbles are below the bisecting line. In relation to the axes scale, and just as a curiosity related to the differences in the citation habits among fields, it is interesting to highlight that the average number of citations in SCIE is approximately twice that in SSCI, and seven times higher than that in AHCI.

[Figure 2 about here]

We define the *OA citation advantage* (OACA) as the average citation of OA articles in relation to non-OA articles. More precisely, denoting by $OAC_i$ the average citation of OA articles in category *i*, and by $NOAC_i$ the average citation of non-OA articles in category *i*, then the OA citation advantage of thematic category *i* can be defined as:

$$OACA_i = \frac{OAC_i - NOAC_i}{NOAC_i} \times 100$$

Therefore, a value of *OACA=p* means that OA articles are cited *p*% more than non-OA articles. Similarly, a negative value of -*p*% means that OA articles are cited *p*% less than non-OA articles.

In order to clarify this definition, consider the example in Table 3. The OA citation advantage of category 1 is about 33%, that is, in category 1 the OA articles are cited 33% more than non-OA articles. Similarly, the OA citation advantage of category 2 is -25%, that is, in category 2 the OA articles are cited a 25% less than non-OA articles.

[Table 3 about here]

Note in Figure 3 that most categories are below 0% in the OA citation advantage, and therefore there is no generalizable OA citation advantage at article level in the Web of Science database.

[Figure 3 about here]

The OA citation advantage by different groups and signs is analysed in Table 4. Focusing on the group of categories in which the OA prevalence is appreciable (over 5%), just one in four categories have positive OA citation advantage (in 24 categories is positive while in 71 is negative). Furthermore, in half of the categories (median), the OA citation advantage is less than 22%, while the OA citation disadvantage is higher than 54%.

[Table 4 about here]

In addition, it should be noted that five out of the seven subject categories with OA prevalence above 15% and positive OA citation advantage, correspond to thematic categories related to Biomedicine (OA prevalence, OA citation advantage): 'Tropical Medicine' (43.4%, 28.43%), 'Primary Health Care' (37.5%, 64.30%), 'Parasitology' (35.2%, 98.16%), 'Mathematical & Computational Biology' (27.6%, 5.18%), 'Folklore' (20.3%, 19.05%), 'Meteorology & Atmospheric Sciences' (18.9%, 24.33%), and 'Genetics & Heredity' (15.5%, 7.98%).

The OA citation advantage by different indexes is analysed in Table 5. In all three indexes the number of categories where the advantage is negative (disadvantage) is more than four times the number where it is positive. The index with a lower proportion of categories with OA citation advantage is SSCI. However, the advantage in half of the categories (median) is over 39% (versus 20% in science journals). In the case of the arts and humanities journals, positive OA citation advantage is observed only in 3 categories (disadvantage in 14), but the magnitude of this advantage is 92% (above 84% of disadvantage)

[Table 5 about here]

*Journal level analysis*

We analyse secondly the OA at journal level aggregating journals by scientific disciplines (Web of Science subject categories). Figure 4 shows the median journal impact factor for the top 36 categories with the highest OA percentages in 2009. Subject categories are sorted in descending order of the OA proportion in that year. Four categories were excluded from the analysis: 'Crystallography', because it had no OA journal with JIF in 2009, and 'Folklore', 'Religion', and 'Medieval & Renaissance Studies', because they do not have JIF (AHCI).

[Figure 4 about here]

This figure shows how the median JIF of every category changed from 2009 to 2014, for both OA and non-OA journals. There seems to be no direct relation between the prevalence in OA and the magnitude of the median JIF. Actually, its correlation is not statistically significant in 2009 (0.18, $p$=0.29) neither in 2014 (0.14, $p$=0.41). There are categories with a high proportion of OA but a low median JIF, and categories with a lower proportion of OA but a higher median JIF. One can also see that most categories –29 out of 36– have a higher median JIF in 2009 for their non-OA journals than for their OA ones. This relation changes only slightly in 2014, as there remain 27 out of 36 categories with a higher median JIF for their non-OA journals than for their OA ones.

Finally, some categories show a nearly parallel trend in their median JIF for both groups, while others cross their trends. Therefore, there is no generalizable OA advantage at journal level in the Web of Science database.

## Conclusions

In relation to the OA prevalence and its changes over time, the percentage of OA articles has increased in the time period 2009-2014 in all three indexes. In SCIE, the average percentage of OA went from 6.34% to 10.17% (increase of 60.4%), in SSCI that percentage went from 3.41% to 4.46% (increase of 30.8%), and in AHCI that percentage went from 3.85% to 4.06% (increase of 5.5%).

In relation to the OA citation advantage, some specific fields have been analysed in the literature. However, this paper is the first multidisciplinary study that includes all scientific disciplines. It also studies the OA effect differentiating at article and journal level. As the main conclusion, there is no generalizable advantage of gold OA in the Web of Science database.

In particular, at both scientific article and journal level, it cannot be concluded that gold OA has increased its impact –neither measured in terms of article citations nor in terms of journal impact factor–. Although there are scientific disciplines where the average impact of an article is higher in the case of OA (36 categories), in most disciplines the opposite happens (173 categories). Something similar happens at journal level, where in those disciplines with the highest OA prevalence, the impact factor of OA journals is mostly lower than that of non-OA journals.

Some considerations can be made for these unexpected results. The main one has to do with the journal visibility. Most OA journals are not at the top of rankings that measure the impact of the journals (e.g. first quartile in the JCR). However, the top of these rankings provides high visibility for the journals. In addition, access through subscription is most widespread amongst journals that are well positioned in those rankings. Thus, gold OA does not guarantee higher visibility in relation to the subscription model.

We do not take into account the influence of the APC costs. The APC of top ranked journals is evidently higher than that of lower ranked journals. For this reason many authors cannot publish in some desired gold open access journals, especially in the top ranked ones.

Finally, as our results are aggregated at subject categories, there may be advantageous and no-advantageous journals/articles which may counterpart their effects, causing some type of misleading similarity.

## Acknowledgements

This research has been supported by Ministerio de Economía, Industria y Competitividad of Spain under the research project ECO2014-59067-P.

Table 1: Prevalence of OA research articles in 2009 and 2014, and average citations in 2009-2014 for articles published in 2009

| | | | 2009 | | 2014 | | Average Citations in 2009-2014 | | | OA Citation Advantage |
|---|---|---|---|---|---|---|---|---|---|---|
| **id** | **WOS Category** | **Index** | **Total Articles** | **% OA** | **Total Articles** | **% OA** | **OA** | **Non-OA** | **Ratio OA/Non-OA** | |
| 1 | Acoustics | SCIE | 3,886 | 0.5% | 4,448 | 5.5% | 2.62 | 8.88 | 0.30 | -70.5% |
| 2 | Agricultural Economics & Policy | SCIE | 663 | 6.2% | 817 | 16.6% | 1.63 | 6.42 | 0.25 | -74.6% |
| 3 | Agricultural Engineering | SCIE | 2,158 | 3.0% | 3,503 | 11.0% | 2.94 | 17.32 | 0.17 | -83.0% |
| 4 | Agriculture, Dairy & Animal Science | SCIE | 6,672 | 19.7% | 6,470 | 12.9% | 3.22 | 7.45 | 0.43 | -56.8% |
| 5 | Agriculture, Multidisciplinary | SCIE | 6,684 | 30.2% | 6,871 | 24.4% | 2.49 | 9.06 | 0.27 | -72.5% |
| 6 | Agronomy | SCIE | 7,365 | 23.1% | 8,335 | 18.0% | 2.66 | 8.36 | 0.32 | -68.2% |
| 7 | Allergy | SCIE | 1,835 | 5.8% | 1,855 | 10.1% | 7.88 | 15.67 | 0.50 | -49.7% |
| 8 | Anatomy & Morphology | SCIE | 1,724 | 10.6% | 1,972 | 17.5% | 2.43 | 8.64 | 0.28 | -71.9% |
| 9 | Andrology | SCIE | 319 | 23.8% | 403 | 18.6% | 7.21 | 8.86 | 0.81 | -18.6% |
| 10 | Anesthesiology | SCIE | 3,479 | 0.7% | 3,287 | 8.8% | 23.36 | 12.10 | 1.93 | 93.1% |
| 11 | Anthropology | SSCI | 2,777 | 8.4% | 3,378 | 13.0% | 1.63 | 6.79 | 0.24 | -76.0% |
| 12 | Archaeology | AHCI | 1,732 | 3.6% | 2,575 | 4.4% | 0.75 | 4.64 | 0.16 | -83.8% |
| 13 | Architecture | AHCI | 5,739 | 1.7% | 1,754 | 4.1% | 0.40 | 4.15 | 0.10 | -90.4% |
| 14 | Area Studies | SSCI | 1,856 | 1.4% | 2,221 | 0.9% | 2.73 | 2.83 | 0.96 | -3.5% |
| 15 | Art | AHCI | 2,334 | 3.3% | 2,376 | 5.0% | 1.58 | 0.48 | 3.29 | 229.2% |
| 16 | Asian Studies | AHCI | 938 | 4.1% | 1,083 | 3.8% | 0.92 | 0.94 | 0.98 | -2.1% |
| 17 | Astronomy & Astrophysics | SCIE | 15,498 | 2.6% | 18,252 | 2.0% | 9.64 | 21.90 | 0.44 | -56.0% |
| 18 | Automation & Control Systems | SCIE | 6,664 | 2.3% | 8,062 | 2.7% | 12.53 | 10.66 | 1.18 | 17.5% |
| 19 | Behavioral Sciences | SCIE | 5,135 | 1.6% | 5,920 | 8.3% | 13.85 | 14.13 | 0.98 | -2.0% |
| 20 | Biochemical Research Methods | SCIE | 12,663 | 8.8% | 14,845 | 11.2% | 19.15 | 16.63 | 1.15 | 15.2% |
| 21 | Biochemistry & Molecular Biology | SCIE | 44,626 | 6.3% | 47,491 | 9.6% | 26.44 | 18.91 | 1.40 | 39.8% |
| 22 | Biodiversity Conservation | SCIE | 3,376 | 6.7% | 4,407 | 9.8% | 4.18 | 12.50 | 0.33 | -66.6% |
| 23 | Biology | SCIE | 84,271 | 8.1% | 10,184 | 27.5% | 17.68 | 16.65 | 1.06 | 6.2% |
| 24 | Biophysics | SCIE | 10,844 | 0.8% | 12,527 | 0.9% | 7.89 | 13.82 | 0.57 | -42.9% |
| 25 | Biotechnology & Applied Microbiology | SCIE | 21,414 | 15.4% | 25,906 | 23.0% | 12.48 | 15.81 | 0.79 | -21.1% |
| 26 | Business | SSCI | 8,188 | 3.2% | 5,535 | 1.6% | 2.53 | 9.63 | 0.26 | -73.7% |
| 27 | Business, Finance | SSCI | 3,392 | 1.1% | 4,306 | 0.3% | 4.63 | 8.92 | 0.52 | -48.1% |
| 28 | Cardiac & Cardiovascular Systems | SCIE | 14,512 | 9.9% | 16,593 | 13.9% | 6.66 | 17.71 | 0.38 | -62.4% |
| 29 | Cell & Tissue Engineering | SCIE | 1,643 | 0.6% | 2,417 | 18.7% | 33.80 | 28.05 | 1.20 | 20.5% |
| 30 | Cell Biology | SCIE | 19,349 | 2.2% | 23,359 | 15.1% | 9.50 | 25.61 | 0.37 | -62.9% |
| 31 | Chemistry, Analytical | SCIE | 17,400 | 4.1% | 21,592 | 6.3% | 7.49 | 12.61 | 0.59 | -40.6% |
| 32 | Chemistry, Applied | SCIE | 18,056 | 1.2% | 13,007 | 2.2% | 3.93 | 16.27 | 0.24 | -75.8% |
| 33 | Chemistry, Inorganic & Nuclear | SCIE | 12,335 | 0.4% | 12,857 | 0.8% | 13.24 | 10.99 | 1.20 | 20.5% |
| 34 | Chemistry, Medicinal | SCIE | 10,656 | 2.9% | 12,534 | 6.0% | 3.80 | 13.12 | 0.29 | -71.0% |
| 35 | Chemistry, Multidisciplinary | SCIE | 49,665 | 5.2% | 57,421 | 5.3% | 4.82 | 20.22 | 0.24 | -76.2% |
| 36 | Chemistry, Organic | SCIE | 19,386 | 3.7% | 20,147 | 8.4% | 8.21 | 13.10 | 0.63 | -37.3% |
| 37 | Chemistry, Physical | SCIE | 41,741 | 0.2% | 54,181 | 0.6% | 11.64 | 17.81 | 0.65 | -34.6% |
| 38 | Classics | AHCI | 707 | 5.1% | 790 | 6.1% | 1.25 | 0.65 | 1.92 | 92.3% |
| 39 | Clinical Neurology | SCIE | 20,404 | 2.8% | 22,659 | 5.4% | 8.40 | 15.28 | 0.55 | -45.0% |
| 40 | Communication | SSCI | 2,446 | 3.4% | 2,940 | 5.7% | 1.55 | 7.39 | 0.21 | -79.0% |
| 41 | Computer Science, Artificial Intelligence | SCIE | 8,621 | 3.6% | 11,727 | 4.4% | 8.59 | 11.60 | 0.74 | -25.9% |
| 42 | Computer Science, Cybernetics | SCIE | 1,120 | 1.4% | 1,377 | 5.9% | 4.81 | 7.74 | 0.62 | -37.9% |
| 43 | Computer Science, Hardware & Architecture | SCIE | 3,740 | 0.0% | 4,876 | 0.0% | -- | 6.11 | -- | -- |
| 44 | Computer Science, Information Systems | SCIE | 8,449 | 3.6% | 12,629 | 6.6% | 2.46 | 8.22 | 0.30 | -70.1% |
| 45 | Computer Science, Interdisciplinary Applications | SCIE | 10,489 | 1.0% | 13,043 | 1.5% | 12.25 | 11.37 | 1.08 | 7.7% |
| 46 | Computer Science, Software Engineering | SCIE | 6,635 | 1.0% | 8,072 | 1.8% | 2.25 | 6.57 | 0.34 | -65.8% |
| 47 | Computer Science, Theory & Methods | SCIE | 5,677 | 0.7% | 7,498 | 1.1% | 0.47 | 6.75 | 0.07 | -93.0% |
| 48 | Construction & Building Technology | SCIE | 3,865 | 1.9% | 6,964 | 2.2% | 1.36 | 7.81 | 0.17 | -82.6% |
| 49 | Criminology & Penology | SSCI | 1,534 | 0.0% | 1,996 | 0.0% | -- | 7.01 | -- | -- |
| 50 | Critical Care Medicine | SCIE | 3,954 | 1.2% | 4,121 | 1.8% | 2.73 | 18.27 | 0.15 | -85.1% |
| 51 | Crystallography | SCIE | 10,121 | 41.0% | 6,705 | 1.2% | 1.47 | 9.74 | 0.15 | -84.9% |
| 52 | Cultural Studies | SSCI-AHCI | 789 | 0.0% | 1,226 | 0.0% | -- | 3.24 | -- | -- |
| 53 | Dance | AHCI | 322 | 0.0% | 352 | 0.0% | -- | 0.21 | -- | -- |
| 54 | Demography | SSCI | 721 | 14.4% | 978 | 17.7% | 4.76 | 7.93 | 0.60 | -40.0% |
| 55 | Dentistry, Oral Surgery & Medicine | SCIE | 7,412 | 4.5% | 8,546 | 6.7% | 5.88 | 9.60 | 0.61 | -38.8% |
| 56 | Dermatology | SCIE | 5,882 | 5.6% | 6,308 | 8.0% | 4.98 | 8.59 | 0.58 | -42.0% |
| 57 | Developmental Biology | SCIE | 3,663 | 3.0% | 3,553 | 3.2% | 15.41 | 19.46 | 0.79 | -20.8% |
| 58 | Ecology | SCIE | 14,243 | 4.1% | 16,546 | 9.4% | 11.36 | 15.61 | 0.73 | -27.2% |
| 59 | Economics | SSCI | 14,823 | 2.6% | 17,904 | 2.9% | 1.57 | 9.16 | 0.17 | -82.9% |
| 60 | Education & Educational Research | SSCI | 7,551 | 7.2% | 9,777 | 9.1% | 4.17 | 6.41 | 0.65 | -34.9% |
| 61 | Education, Scientific Disciplines | SCIE | 2,675 | 11.3% | 3,328 | 16.9% | 6.17 | 6.80 | 0.91 | -9.3% |
| 62 | Education, Special | SSCI | 980 | 0.0% | 1,466 | 0.0% | -- | 9.43 | -- | -- |
| 63 | Electrochemistry | SCIE | 9,175 | 6.6% | 14,512 | 12.3% | 9.84 | 17.48 | 0.56 | -43.7% |
| 64 | Emergency Medicine | SCIE | 2,544 | 8.1% | 3,155 | 9.1% | 2.51 | 7.69 | 0.33 | -67.4% |
| 65 | Endocrinology & Metabolism | SCIE | 12,823 | 4.0% | 14,425 | 8.5% | 8.30 | 18.19 | 0.46 | -54.4% |
| 66 | Energy & Fuels | SCIE | 12,404 | 1.7% | 26,816 | 3.7% | 8.19 | 15.15 | 0.54 | -45.9% |
| 67 | Engineering, Aerospace | SCIE | 2,517 | 0.0% | 3,115 | 0.7% | -- | 3.86 | -- | -- |
| 68 | Engineering, Biomedical | SCIE | 8,371 | 1.1% | 12,016 | 3.6% | 12.96 | 14.40 | 0.90 | -10.0% |
| 69 | Engineering, Chemical | SCIE | 20,599 | 2.4% | 28,566 | 2.1% | 3.63 | 11.18 | 0.32 | -67.5% |
| 70 | Engineering, Civil | SCIE | 11,531 | 1.0% | 15,832 | 2.2% | 1.26 | 10.71 | 0.12 | -88.2% |
| 71 | Engineering, Electrical & Electronic | SCIE | 39,902 | 3.0% | 48,660 | 3.9% | 5.68 | 8.96 | 0.63 | -36.6% |
| 72 | Engineering, Environmental | SCIE | 10,228 | 0.4% | 11,240 | 1.3% | 8.02 | 19.52 | 0.41 | -58.9% |
| 73 | Engineering, Geological | SCIE | 1,920 | 0.0% | 2,959 | 0.0% | -- | 6.87 | -- | -- |
| 74 | Engineering, Industrial | SCIE | 4,294 | 0.0% | 4,702 | 1.0% | -- | 9.15 | -- | -- |
| 75 | Engineering, Manufacturing | SCIE | 4,751 | 0.4% | 5,386 | 0.5% | 5.60 | 8.08 | 0.69 | -30.7% |
| 76 | Engineering, Marine | SCIE | 674 | 0.0% | 822 | 13.4% | -- | 3.09 | -- | -- |
| 77 | Engineering, Mechanical | SCIE | 13,111 | 0.6% | 17,259 | 5.3% | 2.12 | 8.08 | 0.26 | -73.8% |
| 78 | Engineering, Multidisciplinary | SCIE | 29,336 | 3.0% | 11,516 | 23.3% | 2.36 | 9.17 | 0.26 | -74.3% |
| 79 | Engineering, Ocean | SCIE | 923 | 0.0% | 1,252 | 0.0% | -- | 7.16 | -- | -- |
| 80 | Engineering, Petroleum | SCIE | 1,561 | 6.5% | 2,069 | 5.5% | 4.51 | 2.26 | 2.00 | 99.6% |
| 81 | Entomology | SCIE | 5,221 | 6.9% | 5,826 | 13.3% | 3.17 | 6.57 | 0.48 | -51.8% |
| 82 | Environmental Sciences | SCIE | 31,097 | 4.0% | 39,177 | 6.7% | 13.21 | 14.24 | 0.93 | -7.2% |
| 83 | Environmental Studies | SSCI | 4,730 | 1.9% | 6,881 | 11.1% | 14.72 | 10.60 | 1.39 | 38.9% |
| 84 | Ergonomics | SSCI | 979 | 0.0% | 1,459 | 0.0% | -- | 8.46 | -- | -- |
| 85 | Ethics | SSCI | 2,115 | 4.6% | 2,087 | 7.5% | 2.44 | 5.65 | 0.43 | -56.8% |
| 86 | Ethnic Studies | SSCI | 519 | 0.0% | 716 | 0.0% | -- | 5.51 | -- | -- |
| 87 | Evolutionary Biology | SCIE | 4,863 | 8.0% | 5,321 | 12.6% | 18.18 | 17.13 | 1.06 | 6.1% |
| 88 | Family Studies | SSCI | 1,707 | 1.0% | 2,189 | 0.4% | 14.18 | 7.98 | 1.78 | 77.7% |
| 89 | Film, Radio, Television | AHCI | 931 | 2.4% | 1,089 | 13.9% | 0.05 | 1.38 | 0.04 | -96.4% |
| 90 | Fisheries | SCIE | 4,438 | 5.0% | 4,749 | 8.9% | 4.24 | 8.37 | 0.51 | -49.3% |
| 91 | Folklore | AHCI | 276 | 20.3% | 298 | 21.1% | 0.50 | 0.42 | 1.19 | 19.0% |
| 92 | Food Science & Technology | SCIE | 16,196 | 3.6% | 20,366 | 2.5% | 3.07 | 10.18 | 0.30 | -69.8% |
| 93 | Forestry | SCIE | 4,015 | 11.4% | 4,893 | 19.0% | 3.04 | 9.30 | 0.33 | -67.3% |
| 94 | Gastroenterology & Hepatology | SCIE | 9,716 | 10.7% | 11,309 | 21.0% | 9.49 | 16.90 | 0.56 | -43.8% |
| 95 | Genetics & Heredity | SCIE | 15,560 | 15.5% | 19,534 | 22.2% | 24.08 | 22.30 | 1.08 | 8.0% |
| 96 | Geochemistry & Geophysics | SCIE | 8,010 | 3.1% | 10,047 | 4.1% | 3.74 | 12.32 | 0.30 | -69.6% |
| 97 | Geography | SSCI | 5,658 | 2.3% | 3,924 | 5.7% | 6.02 | 11.15 | 0.54 | -46.0% |
| 98 | Geography, Physical | SCIE | 3,397 | 2.2% | 5,471 | 5.2% | 9.99 | 12.55 | 0.80 | -20.4% |
| 99 | Geology | SCIE | 2,112 | 9.0% | 2,380 | 16.9% | 6.89 | 9.56 | 0.72 | -27.9% |
| 100 | Geosciences, Multidisciplinary | SCIE | 15,250 | 9.4% | 21,031 | 11.7% | 12.14 | 11.14 | 1.09 | 9.0% |
| 101 | Geriatrics & Gerontology | SCIE | 3,152 | 1.2% | 4,766 | 17.0% | 3.23 | 13.90 | 0.23 | -76.8% |
| 102 | Gerontology | SSCI | 3,824 | 1.0% | 2,650 | 7.7% | 3.23 | 12.98 | 0.25 | -75.1% |
| 103 | Health Care Sciences & Services | SCIE | 6,023 | 8.9% | 8,042 | 18.2% | 14.49 | 10.95 | 1.32 | 32.3% |
| 104 | Health Policy & Services | SSCI | 3,832 | 8.4% | 5,325 | 12.3% | 12.91 | 11.15 | 1.16 | 15.8% |
| 105 | Hematology | SCIE | 8,876 | 3.1% | 9,403 | 4.7% | 17.76 | 20.55 | 0.86 | -13.6% |
| 106 | History | SSCI-AHCI | 7,606 | 6.2% | 6,632 | 7.2% | 0.47 | 1.68 | 0.28 | -72.0% |
| 107 | History & Philosophy of Science | SCIE-SSCI-AHCI | 1,725 | 8.9% | 1,000 | 0.0% | 0.88 | 3.60 | 0.24 | -75.6% |
| 108 | History of Social Sciences | SSCI | 771 | 0.0% | 2,268 | 7.3% | -- | 2.98 | -- | -- |
| 109 | Horticulture | SCIE | 3,085 | 9.7% | 3,405 | 9.0% | 1.70 | 8.05 | 0.21 | -78.9% |
| 110 | Hospitality, Leisure, Sport & Tourism | SSCI | 1,390 | 0.0% | 2,170 | 0.0% | -- | 7.92 | -- | -- |
| 111 | Humanities, Multidisciplinary | AHCI | 3,290 | 5.9% | 3,271 | 5.4% | 0.18 | 0.81 | 0.22 | -77.8% |
| 112 | Imaging Science & Photographic Technology | SCIE | 2,101 | 3.3% | 3,980 | 1.7% | 1.74 | 14.24 | 0.12 | -87.8% |
| 113 | Immunology | SCIE | 18,087 | 5.0% | 18,620 | 9.6% | 15.43 | 20.65 | 0.75 | -25.3% |
| 114 | Industrial Relations & Labor | SSCI | 754 | 14.9% | 903 | 8.6% | 6.69 | 5.91 | 1.13 | 13.2% |
| 115 | Infectious Diseases | SCIE | 9,824 | 18.3% | 12,866 | 27.4% | 14.18 | 16.61 | 0.85 | -14.6% |
| 116 | Information Science & Library Science | SSCI | 3,079 | 7.3% | 3,766 | 6.8% | 2.17 | 7.34 | 0.30 | -70.4% |
| 117 | Instruments & Instrumentation | SCIE | 10,519 | 5.1% | 14,320 | 9.4% | 7.77 | 8.46 | 0.92 | -8.2% |
| 118 | Integrative & Complementary Medicine | SCIE | 1,629 | 6.7% | 3,215 | 33.7% | 11.37 | 9.70 | 1.17 | 17.2% |
| 119 | International Relations | SSCI | 2,660 | 1.2% | 3,279 | 1.5% | 2.25 | 4.82 | 0.47 | -53.3% |
| 120 | Language & Linguistics | AHCI | 3,943 | 5.7% | 3,930 | 6.2% | 1.33 | 3.73 | 0.36 | -64.3% |
| 121 | Law | SSCI | 3,865 | 3.0% | 4,288 | 2.0% | 2.17 | 3.65 | 0.59 | -40.5% |
| 122 | Limnology | SCIE | 1,771 | 2.1% | 1,950 | 3.7% | 7.70 | 10.74 | 0.72 | -28.3% |
| 123 | Linguistics | SSCI | 4,776 | 5.0% | 4,658 | 5.3% | 1.89 | 4.66 | 0.41 | -59.4% |
| 124 | Literary Reviews | AHCI | 1,515 | 0.0% | 1,995 | 0.0% | -- | 0.07 | -- | -- |
| 125 | Literary Theory & Criticism | AHCI | 487 | 0.0% | 592 | 0.0% | -- | 0.40 | -- | -- |
| 126 | Literature | AHCI | 6,405 | 3.5% | 3,165 | 3.4% | 0.18 | 0.52 | 0.35 | -65.4% |
| 127 | Literature, African, Australian, Canadian | AHCI | 174 | 0.0% | 133 | 0.0% | -- | 0.61 | -- | -- |
| 128 | Literature, American | AHCI | 277 | 0.0% | 332 | 0.0% | -- | 0.74 | -- | -- |
| 129 | Literature, British Isles | AHCI | 301 | 0.0% | 328 | 0.0% | -- | 0.61 | -- | -- |
| 130 | Literature, German, Dutch, Scandinavian | AHCI | 485 | 0.0% | 419 | 0.0% | -- | 0.35 | -- | -- |
| 131 | Literature, Romance | AHCI | 1,559 | 7.4% | 1,687 | 7.2% | 0.16 | 0.17 | 0.94 | -5.9% |
| 132 | Literature, Slavic | AHCI | 362 | 0.0% | 324 | 0.0% | -- | 0.15 | -- | -- |
| 133 | Management | SSCI | 12,980 | 2.0% | 8,495 | 1.7% | 1.91 | 10.43 | 0.18 | -81.7% |
| 134 | Marine & Freshwater Biology | SCIE | 9,057 | 4.8% | 10,321 | 6.0% | 4.07 | 10.61 | 0.38 | -61.6% |
| 135 | Materials Science, Biomaterials | SCIE | 3,811 | 0.7% | 7,409 | 1.8% | 21.92 | 20.58 | 1.07 | 6.5% |
| 136 | Materials Science, Ceramics | SCIE | 4,125 | 9.5% | 5,563 | 6.1% | 2.07 | 7.38 | 0.28 | -72.0% |
| 137 | Materials Science, Characterization & Testing | SCIE | 2,128 | 0.0% | 2,651 | 0.0% | -- | 4.73 | -- | -- |
| 138 | Materials Science, Coatings & Films | SCIE | 5,140 | 0.0% | 6,639 | 0.0% | -- | 10.48 | -- | -- |
| 139 | Materials Science, Composites | SCIE | 2,367 | 0.0% | 3,667 | 0.0% | -- | 10.03 | -- | -- |
| 140 | Materials Science, Multidisciplinary | SCIE | 54,532 | 1.9% | 76,382 | 6.0% | 7.36 | 15.28 | 0.48 | -51.8% |
| 141 | Materials Science, Paper & Wood | SCIE | 1,234 | 9.8% | 2,053 | 31.6% | 5.92 | 5.13 | 1.15 | 15.4% |
| 142 | Materials Science, Textiles | SCIE | 1,501 | 9.2% | 2,382 | 11.3% | 4.59 | 6.37 | 0.72 | -27.9% |
| 143 | Mathematical & Computational Biology | SCIE | 4,538 | 27.6% | 6,277 | 32.8% | 17.87 | 16.99 | 1.05 | 5.2% |
| 144 | Mathematics | SCIE | 42,524 | 4.8% | 25,134 | 12.2% | 3.09 | 5.96 | 0.52 | -48.2% |
| 145 | Mathematics, Applied | SCIE | 21,659 | 6.2% | 25,087 | 13.8% | 3.15 | 6.80 | 0.46 | -53.7% |
| 146 | Mathematics, Interdisciplinary Applications | SCIE | 8,795 | 3.6% | 9,542 | 24.5% | 2.79 | 7.63 | 0.37 | -63.4% |
| 147 | Mechanics | SCIE | 14,125 | 0.6% | 18,781 | 2.7% | 2.73 | 9.00 | 0.30 | -69.7% |
| 148 | Medical Ethics | SCIE | 689 | 12.0% | 820 | 16.6% | 2.34 | 6.62 | 0.35 | -64.7% |
| 149 | Medical Informatics | SCIE | 1,684 | 5.9% | 3,104 | 10.9% | 16.62 | 10.33 | 1.61 | 60.9% |
| 150 | Medical Laboratory Technology | SCIE | 2,602 | 4.2% | 2,808 | 4.6% | 1.24 | 9.92 | 0.13 | -87.5% |
| 151 | Medicine, General & Internal | SCIE | 16,287 | 28.4% | 18,540 | 41.5% | 5.93 | 13.86 | 0.43 | -57.2% |
| 152 | Medicine, Legal | SCIE | 1,280 | 0.0% | 1,778 | 0.0% | -- | 7.19 | -- | -- |
| 153 | Medicine, Research & Experimental | SCIE | 12,146 | 12.3% | 19,921 | 33.2% | 9.12 | 15.37 | 0.59 | -40.7% |
| 154 | Medieval & Renaissance Studies | AHCI | 566 | 9.4% | 678 | 7.4% | 0.28 | 0.74 | 0.38 | -62.2% |
| 155 | Metallurgy & Metallurgical Engineering | SCIE | 16,224 | 1.9% | 15,794 | 3.2% | 1.85 | 7.98 | 0.23 | -76.8% |
| 156 | Meteorology & Atmospheric Sciences | SCIE | 8,901 | 18.9% | 11,955 | 21.9% | 16.81 | 13.52 | 1.24 | 24.3% |
| 157 | Microbiology | SCIE | 34,048 | 13.1% | 17,608 | 14.4% | 15.40 | 17.14 | 0.90 | -10.2% |
| 158 | Microscopy | SCIE | 833 | 0.0% | 988 | 0.0% | -- | 9.50 | -- | -- |
| 159 | Mineralogy | SCIE | 2,060 | 0.6% | 2,585 | 2.1% | 2.75 | 10.01 | 0.27 | -72.5% |
| 160 | Mining & Mineral Processing | SCIE | 2,352 | 5.2% | 2,632 | 6.0% | 1.10 | 7.04 | 0.16 | -84.4% |
| 161 | Multidisciplinary Sciences | SCIE | 21,016 | 31.9% | 52,193 | 73.8% | 17.38 | 33.91 | 0.51 | -48.7% |
| 162 | Music | AHCI | 1,560 | 0.4% | 1,695 | 0.4% | 0.00 | 1.14 | 0.00 | -100.0% |
| 163 | Mycology | SCIE | 1,488 | 0.3% | 1,744 | 0.7% | 23.50 | 9.10 | 2.58 | 158.2% |
| 164 | Nanoscience & Nanotechnology | SCIE | 17,509 | 2.2% | 29,835 | 9.6% | 12.18 | 22.83 | 0.53 | -46.6% |
| 165 | Neuroimaging | SCIE | 1,905 | 0.0% | 2,695 | 6.5% | -- | 22.93 | -- | -- |
| 166 | Neurosciences | SCIE | 28,695 | 4.4% | 32,966 | 12.9% | 11.51 | 19.52 | 0.59 | -41.0% |
| 167 | Nuclear Science & Technology | SCIE | 9,035 | 0.7% | 9,142 | 1.4% | 3.24 | 5.84 | 0.55 | -44.5% |
| 168 | Nursing | SCIE-SSCI | 5,529 | 9.5% | 6,928 | 5.2% | 1.80 | 6.07 | 0.30 | -70.3% |
| 169 | Nutrition & Dietetics | SCIE | 7,810 | 5.5% | 9,717 | 12.9% | 9.41 | 15.99 | 0.59 | -41.2% |
| 170 | Obstetrics & Gynecology | SCIE | 9,703 | 1.8% | 10,425 | 5.3% | 5.09 | 9.64 | 0.53 | -47.2% |
| 171 | Oceanography | SCIE | 5,468 | 7.8% | 6,412 | 9.2% | 8.62 | 11.26 | 0.77 | -23.4% |
| 172 | Oncology | SCIE | 24,368 | 5.5% | 34,892 | 13.8% | 13.95 | 21.70 | 0.64 | -35.7% |
| 173 | Operations Research & Management Science | SCIE | 7,501 | 0.1% | 8,041 | 0.2% | 3.70 | 9.46 | 0.39 | -60.9% |
| 174 | Ophthalmology | SCIE | 7,814 | 12.2% | 7,939 | 15.5% | 8.84 | 11.17 | 0.79 | -20.9% |
| 175 | Optics | SCIE | 20,537 | 13.6% | 26,716 | 17.9% | 16.80 | 9.76 | 1.72 | 72.1% |
| 176 | Ornithology | SCIE | 1,092 | 0.9% | 1,033 | 1.5% | 5.00 | 5.48 | 0.91 | -8.8% |
| 177 | Orthopedics | SCIE | 8,222 | 7.7% | 10,950 | 11.0% | 6.59 | 11.06 | 0.60 | -40.4% |
| 178 | Otorhinolaryngology | SCIE | 4,770 | 1.7% | 5,084 | 4.6% | 5.60 | 6.97 | 0.80 | -19.7% |
| 179 | Paleontology | SCIE | 2,115 | 8.4% | 2,557 | 9.2% | 6.20 | 8.01 | 0.77 | -22.6% |
| 180 | Parasitology | SCIE | 4,060 | 35.2% | 5,793 | 52.2% | 23.70 | 11.96 | 1.98 | 98.2% |
| 181 | Pathology | SCIE | 8,493 | 6.2% | 7,523 | 22.1% | 4.68 | 11.49 | 0.41 | -59.3% |
| 182 | Pediatrics | SCIE | 12,588 | 4.1% | 14,050 | 5.8% | 4.30 | 9.41 | 0.46 | -54.3% |
| 183 | Peripheral Vascular Disease | SCIE | 8,563 | 1.2% | 9,008 | 2.6% | 9.74 | 19.36 | 0.50 | -49.7% |
| 184 | Pharmacology & Pharmacy | SCIE | 26,094 | 6.6% | 31,155 | 8.5% | 4.54 | 12.65 | 0.36 | -64.1% |
| 185 | Philosophy | AHCI | 5,967 | 6.7% | 5,517 | 5.3% | 0.47 | 2.25 | 0.21 | -79.1% |
| 186 | Physics, Applied | SCIE | 44,337 | 1.9% | 54,729 | 4.4% | 10.31 | 11.91 | 0.87 | -13.4% |
| 187 | Physics, Atomic, Molecular & Chemical | SCIE | 13,923 | 0.2% | 16,764 | 1.8% | 6.10 | 12.78 | 0.48 | -52.3% |
| 188 | Physics, Condensed Matter | SCIE | 25,793 | 1.9% | 26,254 | 2.0% | 2.99 | 13.95 | 0.21 | -78.6% |
| 189 | Physics, Fluids & Plasmas | SCIE | 7,694 | 0.0% | 8,988 | 0.0% | -- | 9.96 | -- | -- |
| 190 | Physics, Mathematical | SCIE | 11,567 | 1.2% | 9,890 | 2.2% | 3.94 | 8.87 | 0.44 | -55.6% |
| 191 | Physics, Multidisciplinary | SCIE | 40,335 | 7.1% | 21,869 | 17.5% | 8.91 | 15.77 | 0.56 | -43.5% |
| 192 | Physics, Nuclear | SCIE | 8,006 | 2.1% | 5,935 | 4.2% | 8.02 | 7.62 | 1.05 | 5.2% |
| 193 | Physics, Particles & Fields | SCIE | 9,884 | 1.8% | 11,041 | 6.2% | 7.92 | 12.65 | 0.63 | -37.4% |
| 194 | Physiology | SCIE | 9,646 | 5.5% | 9,193 | 12.9% | 9.44 | 14.11 | 0.67 | -33.1% |
| 195 | Planning & Development | SSCI | 2,088 | 0.0% | 2,918 | 0.0% | -- | 7.51 | -- | -- |
| 196 | Plant Sciences | SCIE | 16,501 | 10.4% | 20,206 | 13.3% | 5.35 | 13.73 | 0.39 | -61.0% |
| 197 | Poetry | AHCI | 188 | 0.0% | 124 | 0.0% | -- | 0.26 | -- | -- |
| 198 | Political Science | SSCI | 5,415 | 2.3% | 6,227 | 1.6% | 1.47 | 5.32 | 0.28 | -72.4% |
| 199 | Polymer Science | SCIE | 14,260 | 1.3% | 17,258 | 2.5% | 5.74 | 12.74 | 0.45 | -54.9% |
| 200 | Primary Health Care | SCIE | 833 | 37.5% | 1,384 | 44.9% | 8.79 | 5.35 | 1.64 | 64.3% |
| 201 | Psychiatry | SCIE-SSCI | 13,611 | 5.8% | 16,479 | 8.1% | 4.72 | 15.73 | 0.30 | -70.0% |
| 202 | Psychology | SCIE | 27,885 | 2.3% | 6,707 | 13.3% | 5.38 | 12.98 | 0.41 | -58.6% |
| 203 | Psychology, Applied | SSCI | 2,802 | 1.7% | 3,531 | 1.5% | 2.02 | 11.47 | 0.18 | -82.4% |
| 204 | Psychology, Biological | SSCI | 1,429 | 0.0% | 1,552 | 0.0% | -- | 12.70 | -- | -- |
| 205 | Psychology, Clinical | SSCI | 5,567 | 0.9% | 7,044 | 1.6% | 6.46 | 12.81 | 0.50 | -49.6% |
| 206 | Psychology, Developmental | SSCI | 3,701 | 0.0% | 4,410 | 0.0% | -- | 15.27 | -- | -- |
| 207 | Psychology, Educational | SSCI | 1,849 | 0.8% | 2,110 | 0.9% | 7.50 | 10.05 | 0.75 | -25.4% |
| 208 | Psychology, Experimental | SSCI | 5,367 | 0.5% | 6,849 | 1.5% | 6.00 | 14.66 | 0.41 | -59.1% |
| 209 | Psychology, Mathematical | SSCI | 664 | 0.0% | 543 | 0.0% | -- | 11.90 | -- | -- |
| 210 | Psychology, Multidisciplinary | SSCI | 5,787 | 7.7% | 8,043 | 20.2% | 3.44 | 11.33 | 0.30 | -69.6% |
| 211 | Psychology, Psychoanalysis | SSCI | 519 | 0.0% | 498 | 0.0% | -- | 2.54 | -- | -- |
| 212 | Psychology, Social | SSCI | 4,193 | 0.5% | 3,565 | 0.0% | 17.00 | 11.34 | 1.50 | 49.9% |
| 213 | Public Administration | SSCI | 1,485 | 1.4% | 1,682 | 3.6% | 1.00 | 5.72 | 0.17 | -82.5% |
| 214 | Public, Environmental & Occupational Health | SSCI | 19,662 | 16.3% | 25,518 | 20.7% | 9.37 | 11.57 | 0.81 | -19.0% |
| 215 | Radiology, Nuclear Medicine & Medical Imaging | SCIE | 14,736 | 2.4% | 17,963 | 7.4% | 7.19 | 13.94 | 0.52 | -48.4% |
| 216 | Rehabilitation | SCIE-SSCI | 5,375 | 5.5% | 7,455 | 11.8% | 6.81 | 9.30 | 0.73 | -26.8% |
| 217 | Religion | AHCI | 2,686 | 11.8% | 3,148 | 10.4% | 0.52 | 1.22 | 0.43 | -57.4% |
| 218 | Remote Sensing | SCIE | 2,178 | 7.7% | 4,507 | 14.5% | 4.30 | 12.86 | 0.33 | -66.6% |
| 219 | Reproductive Biology | SCIE | 4,144 | 6.2% | 4,102 | 7.8% | 8.40 | 13.29 | 0.63 | -36.8% |
| 220 | Respiratory System | SCIE | 6,543 | 4.2% | 7,853 | 5.8% | 9.44 | 15.91 | 0.59 | -40.7% |
| 221 | Rheumatology | SCIE | 3,446 | 6.4% | 4,167 | 14.0% | 8.42 | 16.04 | 0.52 | -47.5% |
| 222 | Robotics | SCIE | 1,204 | 3.2% | 1,633 | 14.1% | 1.72 | 9.49 | 0.18 | -81.9% |
| 223 | Social Issues | SSCI | 1,379 | 4.8% | 1,641 | 5.6% | 0.85 | 5.90 | 0.14 | -85.6% |
| 224 | Social Sciences, Biomedical | SSCI | 2,150 | 3.9% | 2,849 | 4.8% | 2.36 | 11.78 | 0.20 | -80.0% |
| 225 | Social Sciences, Interdisciplinary | SSCI | 4,798 | 6.6% | 4,989 | 11.0% | 2.43 | 7.19 | 0.34 | -66.2% |
| 226 | Social Sciences, Mathematical Methods | SSCI | 1,838 | 1.7% | 2,214 | 0.7% | 2.74 | 9.28 | 0.30 | -70.5% |
| 227 | Social Work | SSCI | 1,696 | 3.2% | 2,121 | 4.6% | 2.31 | 6.15 | 0.38 | -62.4% |
| 228 | Sociology | SSCI | 4,148 | 4.9% | 5,205 | 4.2% | 1.15 | 6.58 | 0.17 | -82.5% |
| 229 | Soil Science | SCIE | 3,538 | 8.6% | 4,237 | 8.5% | 4.04 | 10.43 | 0.39 | -61.3% |
| 230 | Spectroscopy | SCIE | 7,482 | 0.0% | 8,953 | 0.8% | -- | 8.44 | -- | -- |
| 231 | Sport Sciences | SCIE | 6,541 | 5.1% | 8,243 | 6.0% | 5.68 | 11.65 | 0.49 | -51.2% |
| 232 | Statistics & Probability | SCIE | 8,239 | 4.5% | 9,276 | 5.6% | 5.42 | 10.34 | 0.52 | -47.6% |
| 233 | Substance Abuse | SCIE-SSCI | 2,417 | 2.3% | 3,479 | 3.1% | 7.32 | 12.63 | 0.58 | -42.0% |
| 234 | Surgery | SCIE | 35,413 | 3.0% | 31,893 | 4.4% | 5.37 | 10.63 | 0.51 | -49.5% |
| 235 | Telecommunications | SCIE | 10,082 | 7.8% | 12,925 | 10.2% | 6.68 | 8.17 | 0.82 | -18.2% |
| 236 | Theater | AHCI | 475 | 0.0% | 739 | 1.5% | -- | 0.65 | -- | -- |
| 237 | Thermodynamics | SCIE | 6,136 | 1.5% | 9,898 | 8.7% | 3.18 | 9.82 | 0.32 | -67.6% |
| 238 | Toxicology | SCIE | 8,502 | 9.2% | 9,583 | 7.5% | 16.42 | 12.85 | 1.28 | 27.8% |
| 239 | Transplantation | SCIE | 4,541 | 0.0% | 4,194 | 0.0% | -- | 12.91 | -- | -- |
| 240 | Transportation | SSCI | 3,320 | 0.0% | 3,191 | 1.0% | -- | 7.76 | -- | -- |
| 241 | Transportation Science & Technology | SCIE | 2,587 | 0.0% | 3,615 | 0.0% | -- | 7.16 | -- | -- |
| 242 | Tropical Medicine | SCIE | 2,635 | 43.4% | 3,338 | 52.8% | 11.88 | 9.25 | 1.28 | 28.4% |
| 243 | Urban Studies | SSCI | 1,312 | 1.4% | 2,106 | 1.7% | 1.00 | 8.07 | 0.12 | -87.6% |
| 244 | Urology & Nephrology | SCIE | 9,147 | 3.8% | 9,209 | 8.5% | 5.17 | 13.22 | 0.39 | -60.9% |
| 245 | Veterinary Sciences | SCIE | 13,912 | 22.2% | 12,914 | 21.6% | 2.78 | 6.59 | 0.42 | -57.8% |
| 246 | Virology | SCIE | 5,586 | 13.5% | 6,424 | 19.2% | 32.77 | 18.46 | 1.78 | 77.5% |
| 247 | Water Resources | SCIE | 9,254 | 5.4% | 13,017 | 7.7% | 11.57 | 9.91 | 1.17 | 16.8% |
| 248 | Women's Studies | SSCI | 1,342 | 0.0% | 1,488 | 0.0% | -- | 5.93 | -- | -- |
| 249 | Zoology | SCIE | 9,983 | 10.1% | 11,539 | 13.8% | 3.71 | 8.08 | 0.46 | -54.1% |

OA citation advantage: average citation of OA articles in relation to non-OA articles.
Source: Web of Science

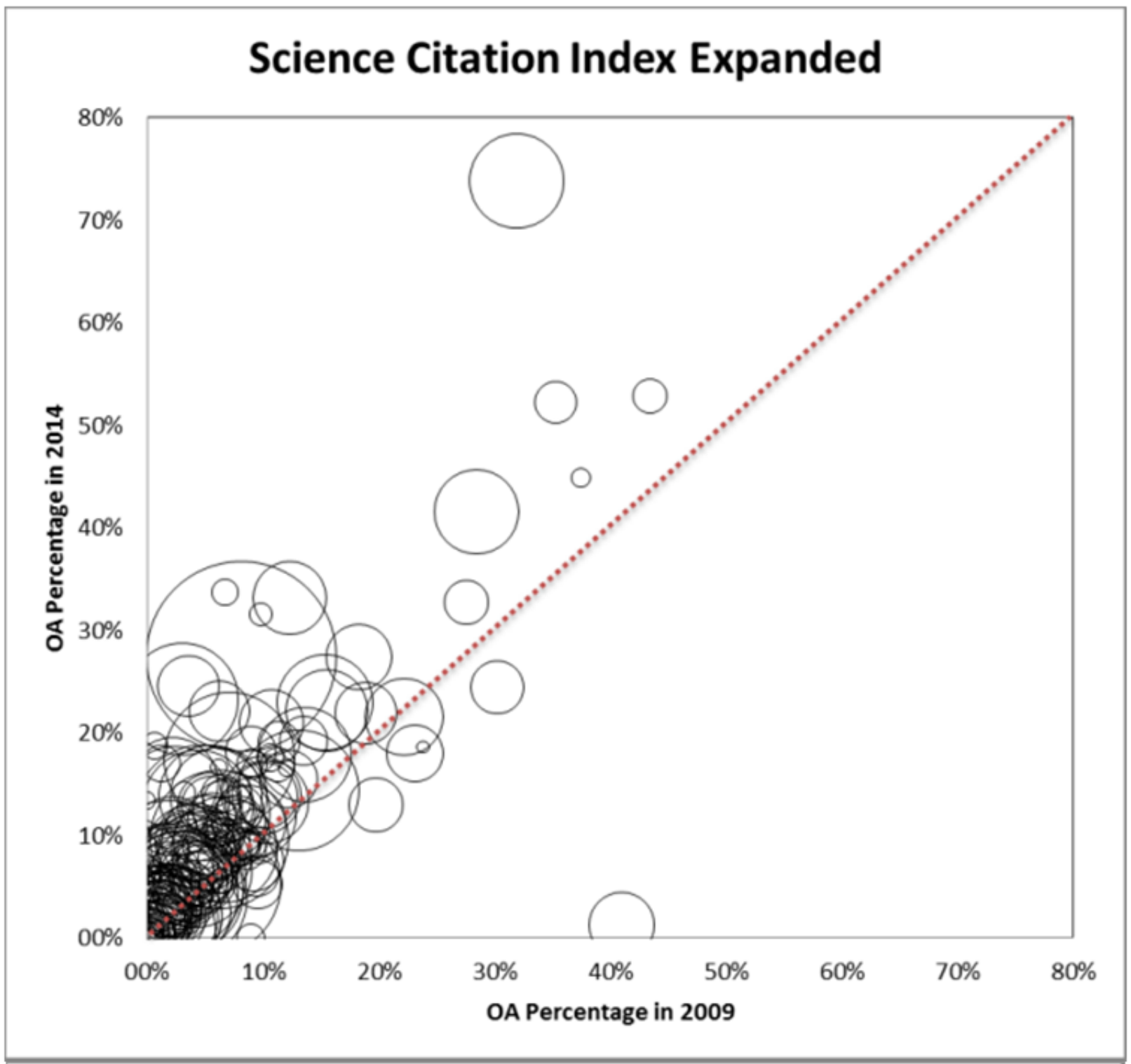
Science Citation Index Expanded
OA Percentage in 2014
OA Percentage in 2009
00%
10%
20%
30%
40%
50%
60%
70%
80%

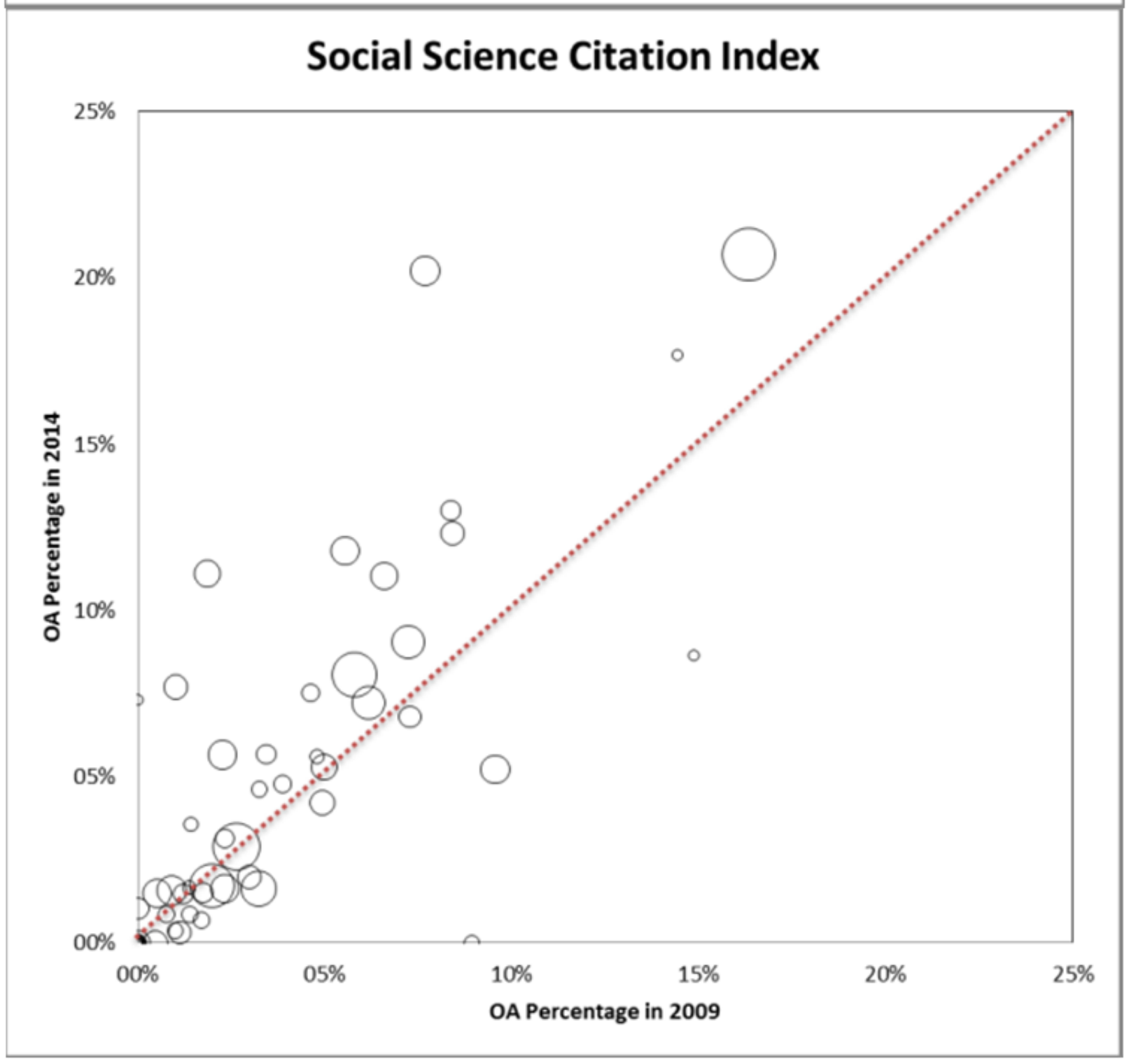
Social Science Citation Index
OA Percentage in 2014
OA Percentage in 2009
00%
05%
10%
15%
20%
25%

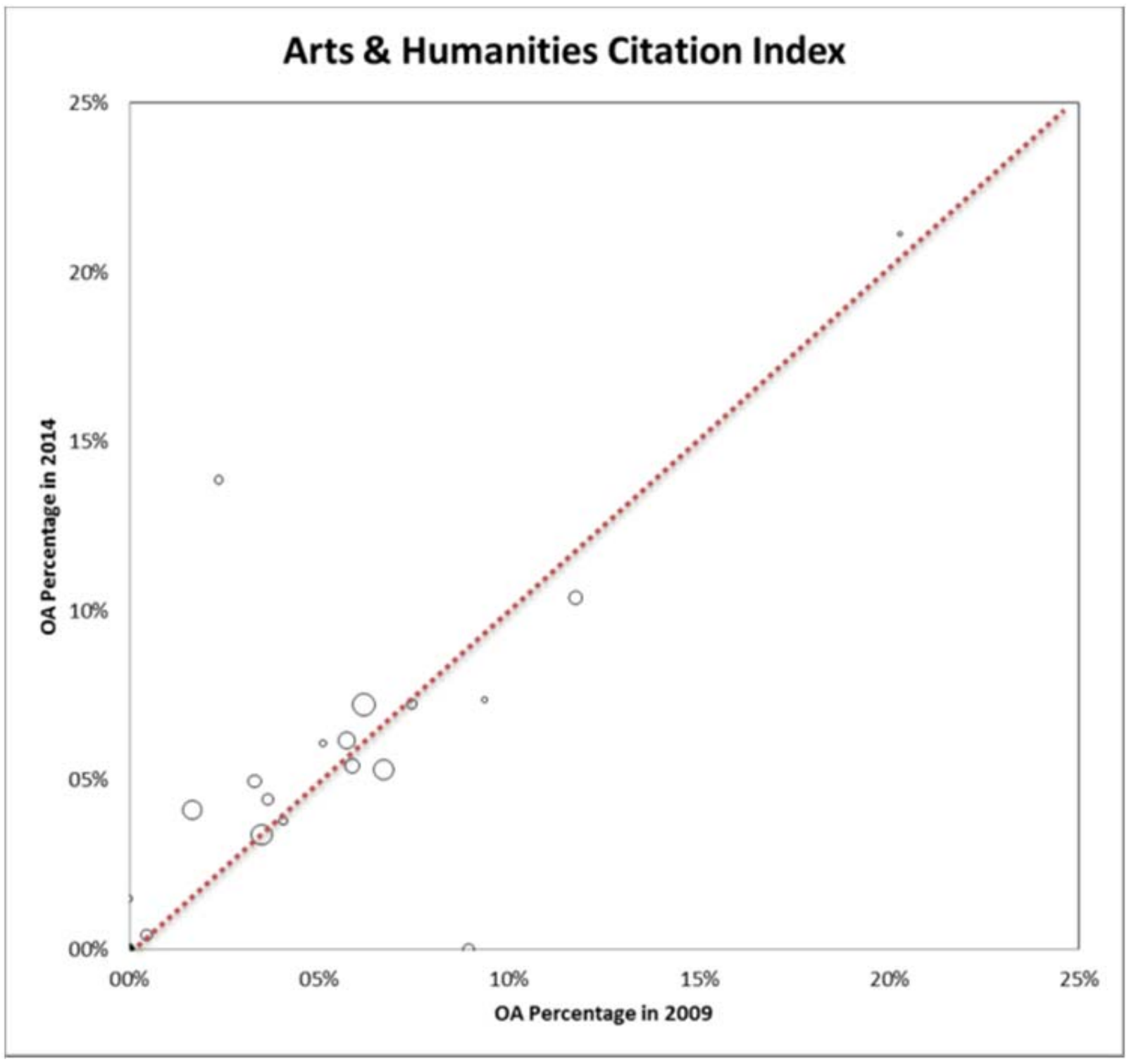

Figure 1: Comparative between OA prevalence in 2009 and 2014 for each category. The bubble size is proportional to the number of research articles in the category

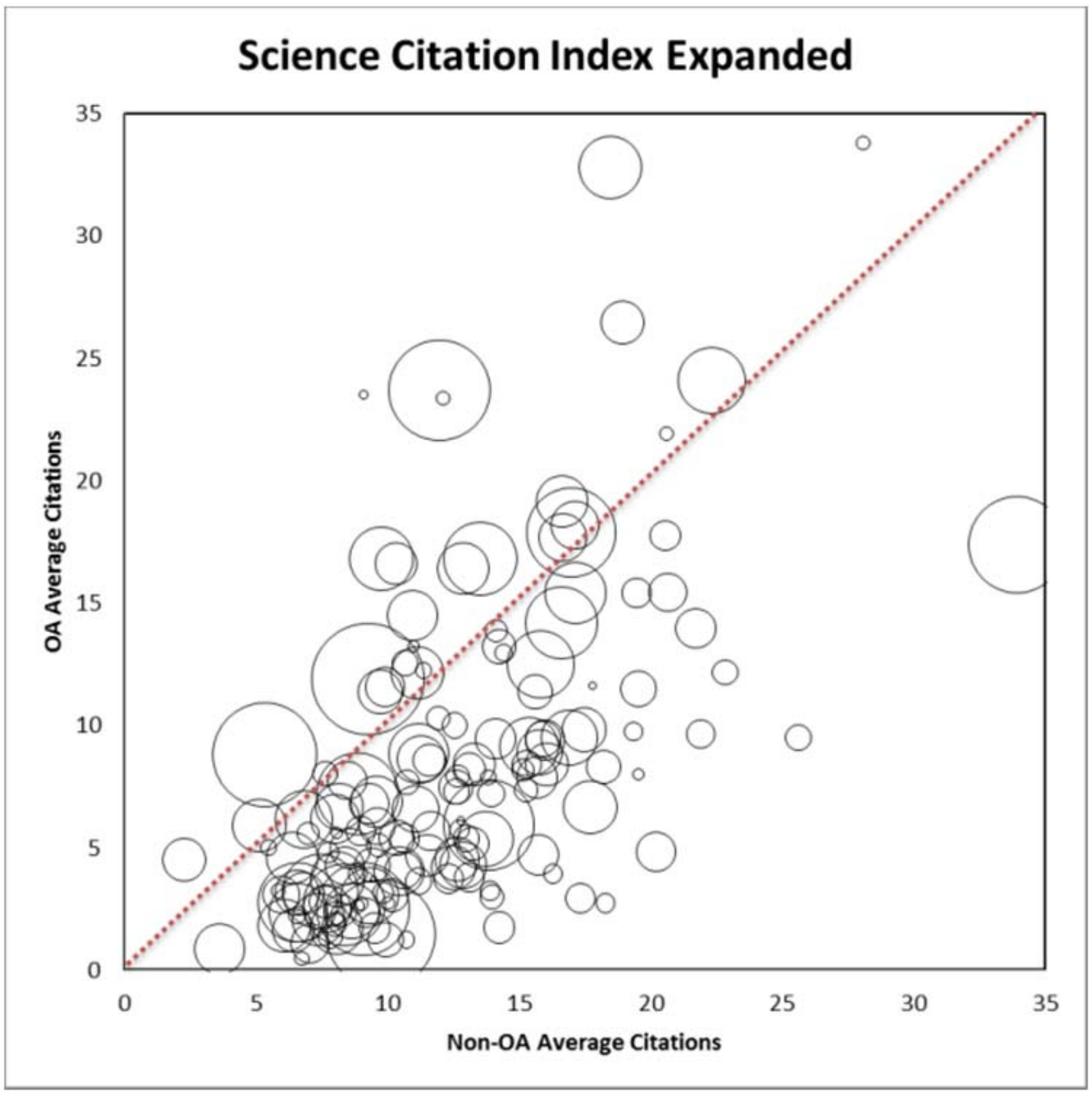

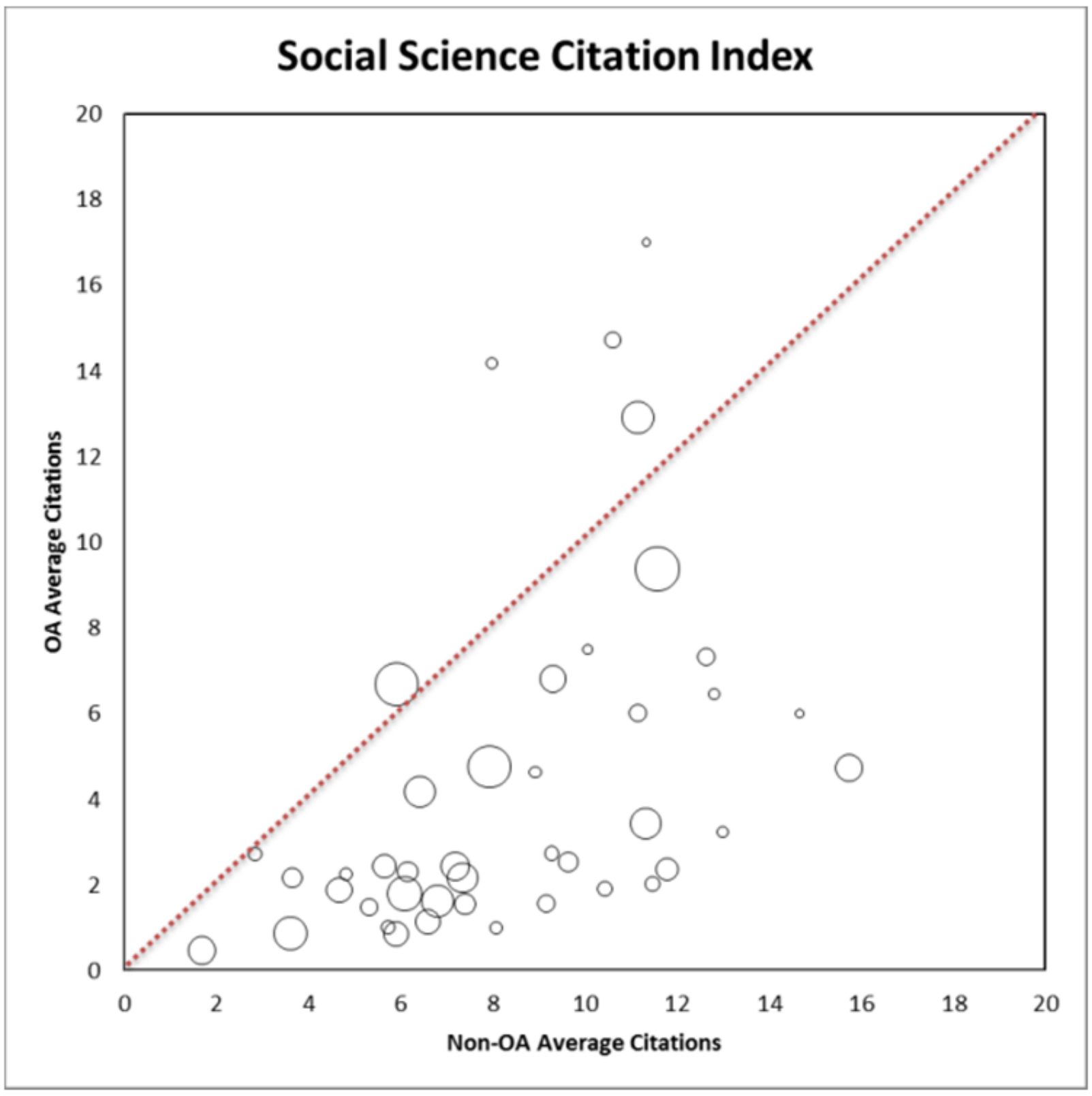

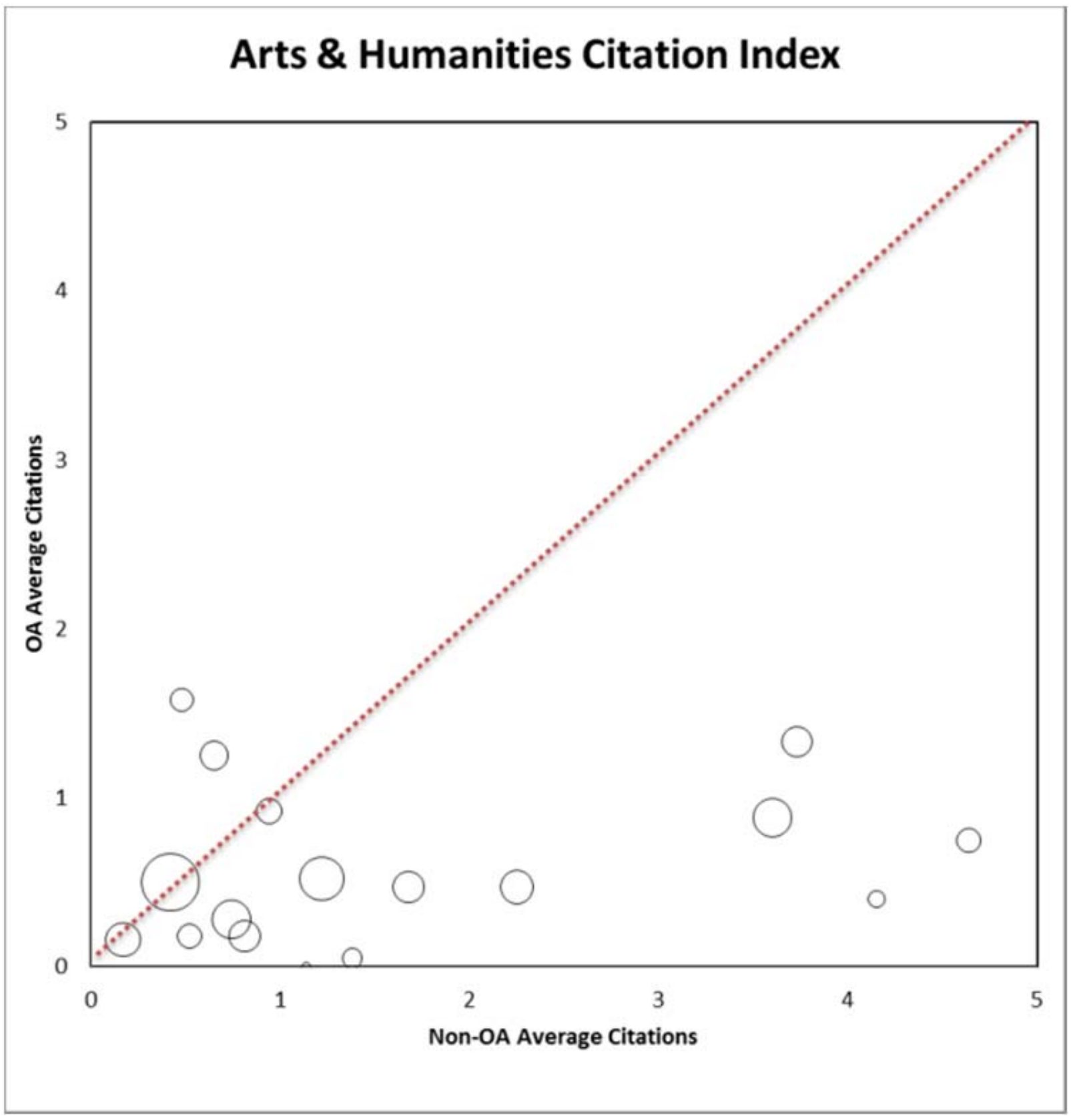

Figure 2: Comparative between OA and non-OA average citations for each category. The bubble size is proportional to the OA prevalence within each category. Most of the bubbles are below de bisecting line.

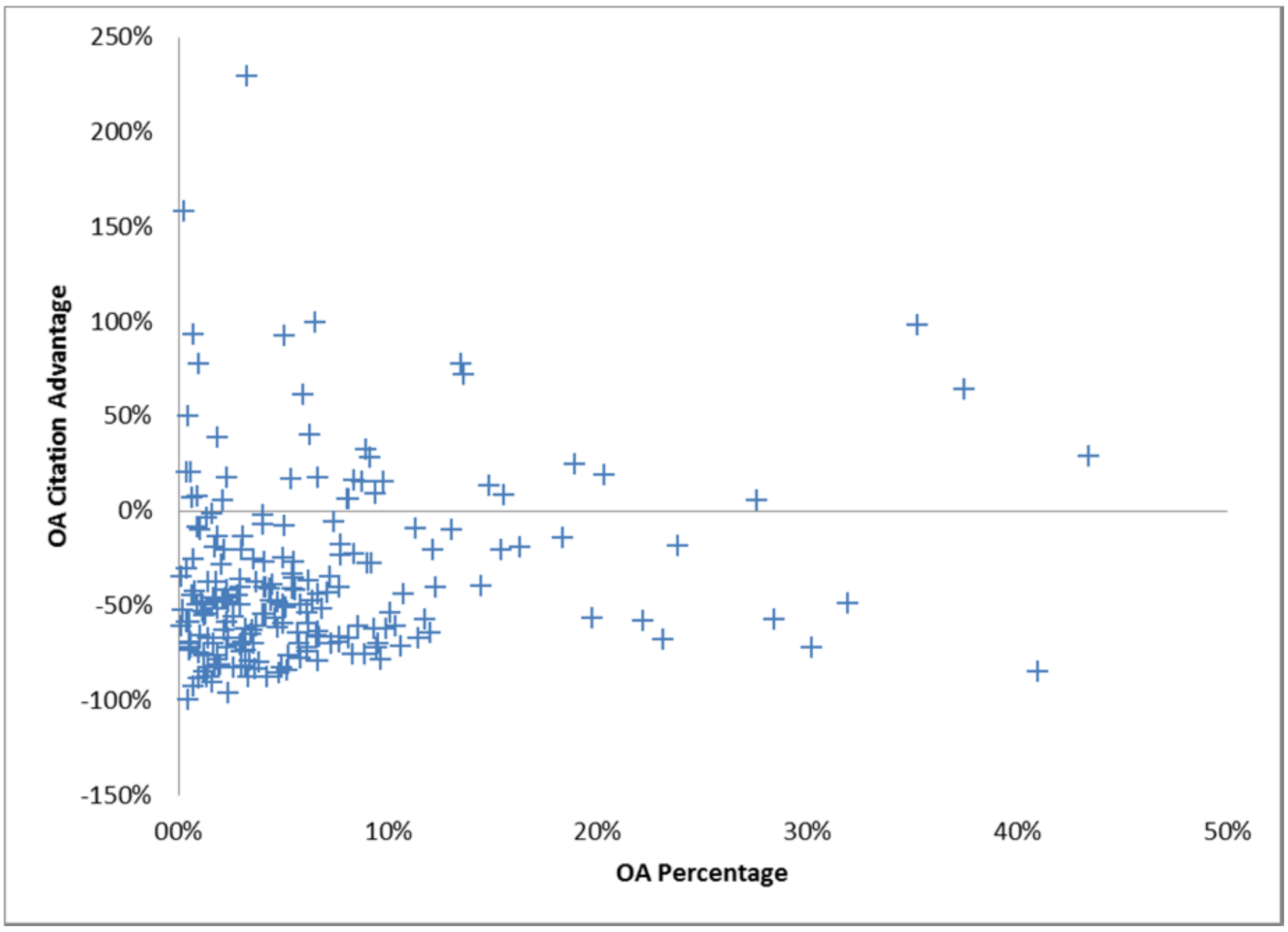

Figure 3: OA citation advantage in relation to OA prevalence for the thematic categories.

Table 2: Descriptive statistics of total articles and OA prevalence

| | | | # Categories | Mean | Median | SD | Min | Max |
|---|---|---|---|---|---|---|---|---|
| 2009 | Total Articles | SCIE | 173 | 10,874.1 | 7,810.0 | 11,863.42 | 319 | 84,271 |
| | | SSCI | 55 | 3,883.4 | 2,718.5 | 3,791.27 | 519 | 19,662 |
| | | AHCI | 26 | 1,728.6 | 934.5 | 1,874.58 | 174 | 6,405 |
| | % OA | SCIE | 173 | 6.3 | 4.1 | 7.89 | 0 | 43.4 |
| | | SSCI | 55 | 3.4 | 2.0 | 3.96 | 0 | 16.3 |
| | | AHCI | 26 | 3.8 | 2.9 | 4.85 | 0 | 20.3 |
| 2014 | Total Articles | SCIE | 176 | 11,142.5 | 7,523.0 | 12,022.40 | 195 | 76,382 |
| | | SSCI | 56 | 4,217.9 | 3,054.5 | 4,577.32 | 211 | 25,518 |
| | | AHCI | 26 | 1,379.1 | 895.0 | 1,310.00 | 124 | 5,517 |
| | % OA | SCIE | 176 | 10.2 | 7.7 | 10.71 | 0 | 73.8 |
| | | SSCI | 56 | 4.5 | 1.9 | 5.22 | 0 | 20.7 |
| | | AHCI | 26 | 4.1 | 3.6 | 5.09 | 0 | 21.1 |

Source: Web of Science

Table 3: Example about the definition of OA citation advantage

| | $OAC_i$ | $NOAC_i$ | $OACA_i = \frac{OAC_i - NOAC_i}{NOAC_i} \times 100$ |
|---|---|---|---|
| Category 1 | 20 | 15 | $OACA_1 = \frac{20-15}{15} \times 100 = \frac{5}{15} \times 100 = \frac{1}{3} \times 100 = 33\%$ |
| Category 2 | 15 | 20 | $OACA_2 = \frac{15-20}{20} \times 100 = -\frac{5}{20} \times 100 = -\frac{1}{4} \times 100 = -25\%$ |

Table 4: Descriptive statistics of the OA citation advantage

| | All categories with OA articles | | Categories over a 5% of OA articles | |
|---|---|---|---|---|
| | Positive | Negative | Positive | Negative |
| # Categories | 36 | 173 | 24 | 71 |
| Mean | 44% | -54% | 36% | -50% |
| Median | 22% | -57% | 22% | -54% |
| Min | 5% | -100% | 5% | -85% |
| Max | 229% | -2% | 100% | -6% |

Source: Web of Science.

Table 5: Descriptive statistics of the OA citation advantage by indexes

| | SCIE | | SSCI | | AHCI | |
|---|---|---|---|---|---|---|
| | Positive | Negative | Positive | Negative | Positive | Negative |
| # Categories | 28 | 129 | 5 | 37 | 3 | 14 |
| Mean | 38% | -51% | 39% | -61% | 113% | -59% |
| Median | 20% | -52% | 39% | -70% | 92% | -84% |
| Min | 5% | -2% | 13% | -3% | 19% | -90% |
| Max | 158% | -93% | 78% | -88% | 229% | -2% |

Source: Web of Science.

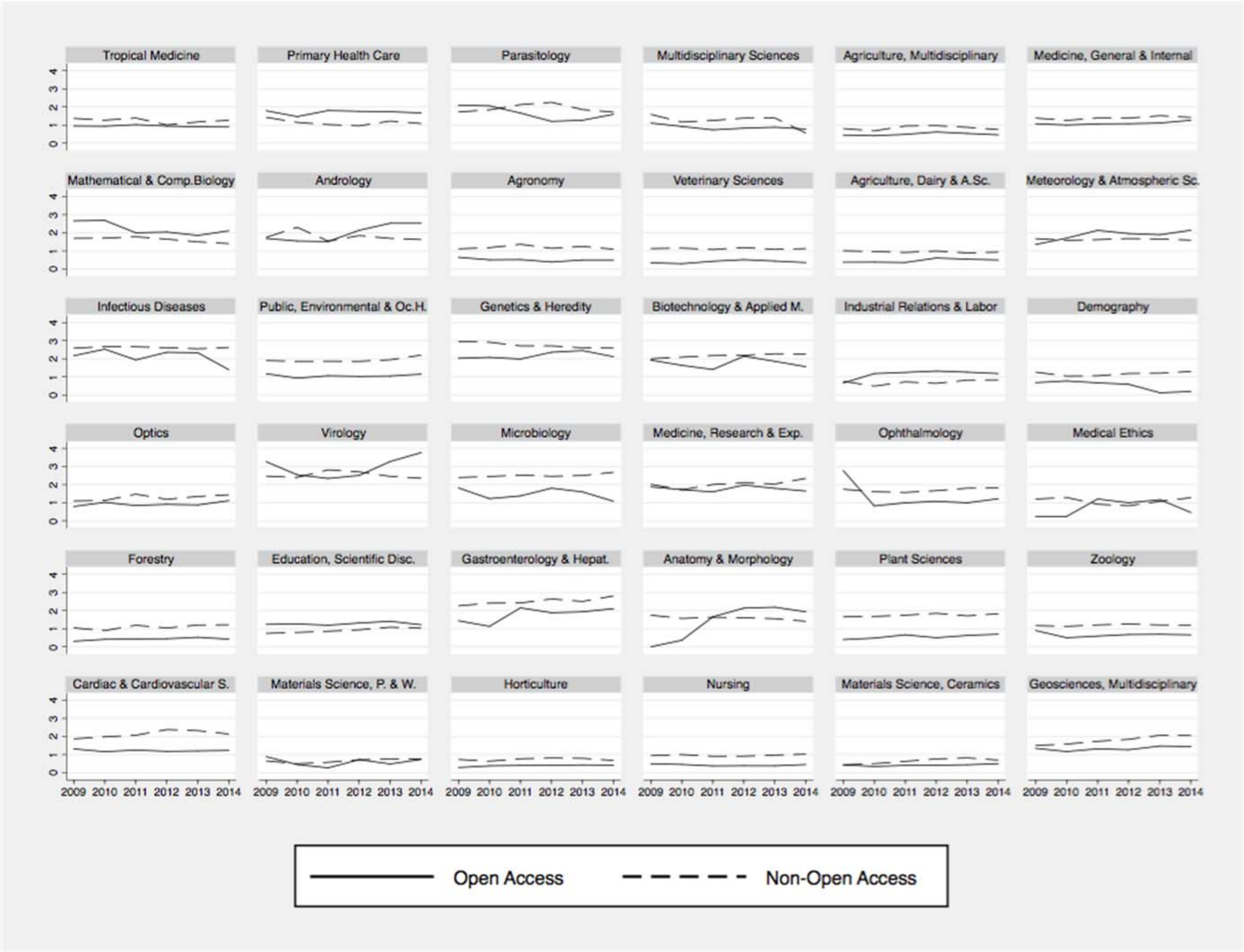

Figure 4: Evolution of the median journal impact factor for the top 36 categories with higher OA percentage in 2009. The category ‘Crystallography’ does not have any OA journal with JIF. Categories ‘Folklore’, ‘Religion’, and ‘Medieval & Renaissance Studies’ do not have JIF (AHCI). Source: Journal Citation Reports.